\documentclass[12pt]{article}

\newcommand{\ww}{\mbox{\tiny $\wedge$}}
\newcommand{\pp}{\partial}
\newcommand{\be}{\begin{equation}}
\newcommand{\ee}{\end{equation}}
\newcommand{\ba}{\begin{eqnarray}}
\newcommand{\ea}{\end{eqnarray}}
\newcommand{\n}{\nonumber\\}

\title{Three-dimensional Noncommutative Gravity }

\author{M.~Ba\~nados, O.~Chand\'{\i}a\\
{\normalsize\it Facultad de F\'\i sica,  P. Universidad Cat\'olica
de Chile}\\ {\normalsize\it Casilla 306, Santiago 22, Chile}\\ ~
\\
N.~Grandi\thanks{CONICET} \, ,
F.A.~Schaposnik\thanks{Associated with CICBA} \, and
 G.A.~Silva$^*$
\\
{\normalsize\it Departamento de F\'\i sica, Universidad Nacional
de La Plata}\\ {\normalsize\it C.C. 67, 1900 La Plata, Argentina}}

\date{\hfill}

\begin{document}

\maketitle

\begin{abstract}

We formulate noncommutative three-dimensional (3d) gravity by making use of its
connection with 3d Chern-Simons theory. In the Euclidean sector, we consider the
particular example of topology $T^2 \times R$ and show that the 3d black hole
solves
the noncommutative equations. We then consider the black hole on a constant U(1)
background and show that the black hole charges (mass and angular momentum) are
modified by the presence of this background.

\end{abstract}
\newpage

\section{Introduction}
Recently, field theories in noncommutative spaces have attracted much
attention, partly in connection with string theory. More
specifically, it has been
shown that noncommutative $U(N)$ gauge theory emerges
in a certain low energy
limit of a system of $Dp$ branes in a constant Neveu-Schwarz $B$ field
background \cite{Connes}-\cite{SW}.
In general,  gauge theories can be  formulated
 in noncommutative spaces starting from  Lagrangians written in
terms of ordinary fields multiplied using the Moyal $*$ product. It should
be noted that consistency requires that the gauge group has to be
$U(N)$ (or certain subgroups of $U(N)$ \cite{armoni}-\cite{Jurco}).

It is
then natural to analyse whether noncommutative extensions can be
also constructed for  gravity. There have been several
investigations on this issue, that basically  start by gauging,
instead of the $SO(d)$ Lorentz group, the
$U(1,d-1)$ \cite{ming}-\cite{Chamseddine3} (or some orthogonal
and symplectic subalgebras of
unitary   groups \cite{armoni}-\cite{Jurco}) and
then define the theory in terms of vielbeins and spin connection
to be multiplied using
the $*$ product.

It is well-known that in three-dimensional space-time, (ordinary) gravity can
be formulated as a Chern-Simons theory \cite{Achucarro},\cite{WittenG}.
Many aspects,  both at the classical and quantum levels, have been understood
using  this connection since, through field redefinitions, it simplifies
the equations and introduces a rich mathematical structure.  The
construction of a black hole in $2+1$ space-time with negative cosmological
constant (the so called BTZ blackhole \cite{BTZ}-\cite{BTZH}) also enhanced
the interest in 3d gravity, particularly in view of the role it plays in  string
theory\cite{Peet}.

The goal of this work is to use the Chern-Simons formulation of
three-dimen\-sion\-al
(3d) gravity to give a definition for 3d noncommutative gravity.
We will take profit on the fact that many classical and quantum aspects of
noncommutative Chern-Simons theory are well understood \cite{Bichl}-\cite{Bak}, to
define the noncommutative 3d gravity action in terms of the corresponding
noncommuta\-tive Chern-Simons action (NCCS)\footnote{There is another kind of
noncommutative field theories, namely the so called q-deformed theories. In this
context, a q-deformed  3d gravity theory has been discussed using the CS
connection \cite{Bimonte}.}.

The paper is organized as follows. We start by
describing in section 2  the NCCS theory  for the group $GL(2,C)$, the one that
will be relevant for the formulation of noncommutative 3d gravity. Then, in
section 3 we establish the connection between gauge fields and gravitational
variables (triad and spin connection) so that the noncommutative ``Einstein
equations'', and their corresponding action,  can be obtained.  We also work out
the metric formulation of the equations. In section 4 we study gravitational
solutions  for the particular topology $M_3=T^2\times \Re$. After showing the
chiral character of these solutions, we construct the corresponding metric and
explore its conformal properties and relate it to the corresponding commutative
solutions. In section 5 we couple the chiral solution to a constant Abelian field
and discuss how noncommutative effects determine the properties of the resulting
blackhole solution.

\section{Noncommutative Chern-Simons theory}

Noncommutative Chern-Simons theory can be defined
by the equations of motion,
\begin{equation}
\pp_\mu  {\cal A}_\nu - \pp_\nu  {\cal A}_\mu + {\cal A}_\mu * {\cal A}_\nu -
 {\cal A}_\nu *
{\cal A}_\mu = 0,
\label{F=0}
\end{equation}
which are invariant under the noncommutative gauge transformations
\begin{equation}
{\cal A}_\mu' = U^{-1} * {\cal A}_\mu * U + U^{-1} * \pp_\mu U.
\label{deltaA}
\end{equation}
Here the $*$ product of two functions $f(x)$ and $g(x)$ is defined as
\begin{equation}
\left.(f*g)(x)  = \exp\left(\frac{i}{2} \theta_{\mu\nu} \partial_{x_\mu}
\partial_{y_\nu}
\right)
f(x)g(y)\right\vert_{y=x}
\label{1}
\end{equation}
with $\theta_{\mu\nu}$ a constant antisymmetric matrix.

The Seiberg-Witten \cite{SW} map provides a powerful method to find
solutions to (\ref{F=0}).  In fact, the key property of this map is
that $\hat {\cal A} +
\hat \delta \hat {\cal A} = \hat {\cal A}({\cal A}
+ \delta {\cal A})$ (with $ {\cal A}$ and $\hat {\cal A}$ gauge fields
for spaces with different values of $\theta_{\mu\nu}$). Thus, if ${\cal A}$ is a
solution to the
commutative equation $d{\cal A}+ {\cal A} {\cal A}=0$, it follows that
$\hat {\cal A}$ is a solution to the
noncommutative equation.

Euclidean gravity, which will be our main interest here, can be formulated as a
Chern-Simons theory for the group $SL(2,C)$. It is well known, however, that in
the
noncommutative case this group is not closed with respect to the Moyal
product and
thus we are forced to consider
$GL(2,C)$.  The gauge field ${\cal A}\in GL(2,C)$ can be expanded in the basis
$\{J_a,i\}$,
\begin{equation}
{\cal A}_\mu = A_\mu^a \, J_a + b_\mu\, i
\label{A}
\end{equation}
where $J_1= (i/2)\sigma_1,
J_2= -(i/2)\sigma_2, J_3= (i/2)\sigma_3$
are anti-Hermitian ($\sigma_a$ are the Pauli matrices).
Since $A^a$ and $b$ are complex, we define a second field
\begin{equation}
\bar {\cal A}_\mu = \bar A_\mu^a \, J_a + \bar b_\mu\, i
\label{Abar}
\end{equation}
which satisfies the Chern-Simons equations as well. It is conventional
to use the
same basis $\{J_a,i\}$ for both fields and thus $\bar
{\cal A}_\mu$ is not the complex
conjugate of ${\cal A}_\mu$.

The Abelian field $b$ can be set equal to zero in the commutative case because it
decouples from $A^a$. This is no longer true in the noncommutative theory,
although
solutions with $b=0$ do exist.

The full set of equations for ${\cal A}$ is,
\begin{eqnarray}
F^a[A] &=& -i (A^a * b + b * A^a) \nonumber\\
db &=&-i( b*b + (1/4)  A^a * A_a)
\label{eqs}
\end{eqnarray}
with $F^a[A] = dA^a + (1/2) \epsilon^a_{\ bc} A^b * A^c $.
The right hand side terms are zero at $\theta=0$ showing that $A^a$ and $b$ are
decoupled in the commutative limit. For future reference, we mention that ``flat"
solutions with $F^a=0$
exist provided,
\begin{equation}
A^a * b + b * A^a =0.
\label{cond}
\end{equation}
Analogous equations can be written for $\bar {\cal A}$.

\section{Three-dimensional noncommutative gravity}

\subsection{Connection representation}

Consider a GL(2,C) gauge field ${\cal A}$, satisfying two copies of (\ref{F=0})
\begin{eqnarray}
d{\cal A} + {\cal A}*{\cal A} &=& 0, \label{CSA} \\
d\bar {\cal A} + \bar {\cal A} * \bar {\cal A} &=& 0 \label{CSAbar}.
\end{eqnarray}
(Here, the wedge symbol has
been omitted). Now we define the combinations
\begin{eqnarray}
{e} &=& {l \over 2i} ({\cal A}- \bar {\cal A}) \nonumber \\
w &=& {1 \over 2} ({\cal A} + \bar {\cal A}) \label{ewA}
\end{eqnarray}
where $e=e^aJ_a + e^4\, i$ and $w=w^aJ_a +w^4 \, i$.
These relations are the natural noncommutative generalization of
\begin{eqnarray}
e^a= {l \over 2i} (A^a- \bar A^a), \nonumber\\
w^a= {1 \over 2} (A^a+ \bar A^a),
\label{relac}
\end{eqnarray}
  Adding and subtracting the
Chern-Simons equations, it is direct to prove that $e$ and $w$ satisfy the
noncommutative ``Einstein Equations",
\begin{eqnarray}
dw + w*w - {1 \over l^2} e*e &=& 0, \label{ncee1} \\
de + w*e + e*w &=& 0 \label{ncee2}
\end{eqnarray}
These equations can be derived from the noncommutative ``Einstein-Hilbert"
action,
\begin{equation}
I[e,w] = \int \mbox{Tr} \left( R * e - {1 \over 3l^2} e*e*e \right),
\label{IEH}
\end{equation}
where  $R = dw + w*w$. The variation with respect to the triad yields
(\ref{ncee1})
while  the variation with respect to $w$ yields the noncommutative torsion
condition (\ref{ncee2}). In deriving the eqs. of motion
from eq.(\ref{IEH}) one has to take into account surface terms which
arise in handling Moyal products (and are absent in the ordinary
commutative case). This terms vanish for the choice of $\theta_{\mu\nu}$
that will be done below (see section 4).

Despite the similarities between the action (\ref{IEH}) and the usual
 Einstein-
Hilbert action, it should be kept in mind that, in the former,
the Abelian fields $b$
and $\bar b$
are coupled to $e^a$ and $w^a$ in a non-trivial way.
The full action (\ref{IEH})
depends on all fields,
\begin{equation}
I = I[e^a, w^a, b, \bar b].
\label{Iew}
\end{equation}
The couplings between $b$ and the gravitational variables are proportional
to $\theta$.  We define noncommutative three-dimensional gravity by this action.

If we set the Abelian fields equal to zero, Eqns. (\ref{ncee1}) and (\ref{ncee2})
become,
\begin{eqnarray}
R^a - {1 \over  l^2}\epsilon^a_{\ bc} e^b * e^c &=& 0,
\label{ncee3} \\ de^a  +(1/2) \epsilon^a_{\ bc} w^b*e^c + (1/2)
\epsilon^a_{\ bc} e^b * w^c &=& 0. \label{ncee4}
\end{eqnarray}
where $R^a = dw^a +(1/2) \epsilon^a_{\ bc} w^b * w^c$.  The first equation can be
regarded as a noncommutative constant curvature condition, written in terms of
connections.  The second equation is the analogous to a torsion condition. This
equation, however, does not imply that the affine  connection is symmetric.

The equations (\ref{ncee3}) and (\ref{ncee4}) are valid provided the equations for
the Abelian field are satisfied with $b=\bar b=0$.  This implies,
\begin{eqnarray}
{1 \over l^{2}} e^a * e_a - w^a*w_a &=& 0 \\ e^a * w_a + w_a * e^a
&=& 0  \label{constraints}
\end{eqnarray}
(which are identically satisfied at $\theta=0$).  We shall display below explicit
solutions fulfilling these conditions.

\subsection{Metric representation}

Eqns. (\ref{ncee3}) and (\ref{ncee4}) have the same form of Einstein equations in
the triad formalism, where all products of functions have been replaced by
the $*$
 product. It is now natural to ask whether there exists a metric formulation
for
them.

We shall assume that the constraints (\ref{constraints}) are satisfied and try to
write (\ref{ncee3}) in terms of the metric and affine  connection. (See
\cite{Moffat}-\cite{Chamseddine2} for other approaches to this problem in four
dimensions.)

We define the metric and affine  connection as\footnote{The definition
of the affine
connection can be motivated by the gauge invariance of the action.
Under gauge
transformations the spin connection transforms as
$w \rightarrow w'= U^{-1} * w * U
+ U^{-1} * dU$.  Let $w'=\Gamma^{\rho}_{\ \lambda
\sigma}$ be the connection in a coordinate basis related
to the tangent basis via
the matrix $U=e^a_{\ \mu}$.
The new connection $\Gamma^{\rho}_{\ \lambda \sigma}$
becomes (\ref{Gamma}).  This equation can also be expressed
as $\pp_\rho e^a_{\
\lambda} +  \epsilon^a_{\ bc}  w^b_{\ \rho} *
e^c_{\ \lambda} -  e^a_{\ \mu} *\Gamma^\mu_{\ \lambda\rho} = 0$,
i.e., the full covariant derivative of $e^a_{\ \mu}$ is zero.},
\begin{eqnarray}
g_{\mu\nu} &=& e^a_{\ \mu} * e^b_{\nu}\, \eta_{ab}, \label{g} \\
\Gamma^\mu_{\ \lambda\rho} &=&  \epsilon_{abc}
e^{\mu a} * w^b_{\ \rho} * e^c_{\ \lambda}+ e^\mu_{\ a} * \pp_\rho e^a_{\
\lambda}.
\label{Gamma}
\end{eqnarray}
In other words, $g_{\mu\nu}$ and $\Gamma^\rho_{\ \mu\nu}$ represent, as usual, the
metric and connection in the coordinate basis.  Given $e^a$ and $w^a$, the above
formulae completely determines $g$ and $\Gamma$. If $e^a$ and $w^a$ satisfy the
Chern-Simons equations, we would like to find the differential equation satisfied
by
$g$ and $\Gamma$.

The curvature in the coordinate basis is,
\begin{equation}
R^\mu_{\ \nu} = d\Gamma^\mu_{\ \nu} + \Gamma^\mu_{\ \sigma} * \Gamma^\sigma_{\
\nu} \ \   \ \  (\Gamma^\mu_{\ \nu} =\Gamma^\mu_{\ \nu\sigma}
dx^\sigma),
\label{Rmunu}
\end{equation}
and it is related to $R^a$ by the formula,
\begin{equation}
 R^\mu_{\ \nu}  = \epsilon_{abc}\, e^{\mu a} * R^b * e^{c}_{\ \nu}.
\label{Ra}
\end{equation}
This follows by direct replacement of (\ref{Gamma}) into (\ref{Rmunu}), and it
expresses the fact that the curvature is a tensor.  Since $R^a$
satisfies (\ref{ncee3}) we find the ``Einstein" equation,
\begin{equation}
R^\mu_{\ \nu \ \alpha\beta}=-{1 \over l^2}( \delta^\mu_{\alpha} g_{\beta\nu} -
\delta^\mu_{\beta} g_{\alpha\nu}) + E^\mu_{\ \nu \ \alpha\beta}
\label{EE}
\end{equation}
where $g_{\mu\nu}$ is defined in (\ref{g}), and
\begin{equation}
E^\mu_{\ \nu \ \alpha\beta} ={1 \over 2l^{2}} e^\mu_{\ a} * ( e^a_{\ [\alpha} *
e^b_{\ \beta]} - e^b_{\ [\beta}  * e^a_{\ \alpha]} ) * e_{b \nu } .
\label{E}
\end{equation}

The first term in (\ref{EE}) is the usual contribution from the cosmological
constant to the Einstein equations.  Recall, however, that in this theory the
metric
is not symmetric.  The second term ($E$) is a purely noncommutative
effect, depending on the commutator of triads with repect to the Moyal product,
and
cannot be expressed in terms of the metric only.

To summarise, given $e^a$ and $w^a$ satisfying the Chern-Simons equations of
motion
then the metric (\ref{g}) and affine  connection (\ref{Gamma}) satisfy the
``Einstein"
equation (\ref{EE}). We shall exhibit below a family of solutions satisfying these
equations.

\section{Solutions}

Before discussing the gravitational solutions, we shall make some general remarks
on
the solutions to the Chern-Simons equations.

All solutions considered here live on the topology $M_3=T^2\times \Re$.
The local coordinates on $T^2$ are $\{z,\bar z\}$ and $\rho \in \Re$.
The components
of the gauge field are then $ {\cal A}_\mu = \{ {\cal A}_z,
{\cal A}_{\bar z},  {\cal A}_\rho\}$.
We shall take $\theta_{\rho z} = \theta_{\rho \bar z} =0$
while  the noncommutative coordinates satisfy,
\begin{equation}
[z,\bar z]=\theta.
\label{zz}
\end{equation}
This means that, to first order in $\theta$,
\begin{equation}
f * g = f \, g + {\theta \over 2} (\pp f \bar \pp g - \bar\pp f\pp g) + {\cal O}(
\theta^2)
\label{f*g}
\end{equation}
with $\pp=\pp/\pp z$, $\bar \pp=\pp/\pp\bar z$. In particular, we find the Moyal
representation of (\ref{zz}),  $z*\bar z -\bar z*z= \theta$.

The choice of manifold $M_3$ and non-trivial component of $\theta_{\mu\nu}$
ensures that when varying the CS action one can use the cyclic property of
the $*$ product without worring about surface terms.
The boundary condition ${\cal A}_{\bar z} = 0$  is required in order to have
well defined functional derivatives of the CS action.

It should be clear that the 3d black hole \cite{BTZ}-\cite{BTZH}
 is a solution to the full noncommutative
 equations simply because this field has two Killing
vectors, $\pp_z$ and $\pp_{\bar z}$,  which effectively reduce the Moyal
product to the usual one.

In order to explore the noncommutative structure, we need to look at more
general
solutions.  We shall start by looking at solutions to the noncommutative
Chern-Simons equations.

\subsection{The chiral solution}
\label{Chiral}

Let us rewrite the first of eqns.(\ref{eqs}) in the form
\ba
&&
F^a_{\rho z}\left[A\right]+
i[b_\rho,A_z^a]  +
i[A_\rho^a,b_z]=0
\n
&&
F^a_{\rho \bar z}\left[A\right]+
i[b_\rho,A_{\bar z}^a]  +
i[A_\rho^a,b_{\bar z}]=0
\n
&&
F^a_{z{\bar z}}\left[A\right]+
i[b_z,A_{\bar z}^a]  +
i[A_z^a,b_{\bar z}]=0
\label{hola}
\ea
where $[A,b] = A*b - b*A$.
Now, fixing the gauge to
\be
A_\rho=iJ^3,\, \;\;\; b_\rho=0
\label{gauge}
\ee
the first two equations (\ref{hola}) become
\ba
&&
\partial_\rho A^a_z  + i\delta^{3}_{\;\, b}\varepsilon^{ab}_{\;\;\;c}{A_z^c}
=0
\n
&&
\partial_\rho A^a_{\bar z}  + i  \delta^{3}_{\;\, b}
\varepsilon^{ab}_{\;\;\;c}
{A_{\bar z}^c}
=0
\ea
with solution
\ba
&&A_z = d^{-1}\tilde A_z(z,\bar z) d
\n
&&A_{\bar z}=d^{-1}\tilde A_{\bar z}(z,\bar z)d
\ea
where
\be
d=e^{i\rho J^3}
\ee
Now, the boundary condition  $A_{\bar z}|_{\partial M}=0$
implies $\tilde A_{\bar z}
=0$, this resulting in
$
A_{\bar z} =0$.
Finally, replacing this solution in the last equation in (\ref{hola}),
we obtain
\ba
 &&
\partial_{\bar z} A^a_{ z} +
i[b_{\bar z},A_z^a]= D_{\bar z}[b]A^a_{ z}=0
\label{u}
\ea

Let us now study the  last equation in (\ref{eqs})
\ba
&&
\partial_\rho b_z - \partial_z b_\rho + i[b_\rho, b_z] +
\frac{i}{4}[A_\rho^a,A_{za}]=0
\n
&&
\partial_\rho b_{\bar z} - \partial_{\bar z} b_\rho + i[b_\rho, b_{\bar z}] +
i
\frac{i}{4}[A_\rho^a,A_{\bar z a}]=0
\n
&&
\partial_z b_{\bar z} - \partial_{\bar z} b_z + i[b_z, b_{\bar z}] + i
\frac{i}{4}[A_z^a,A_{\bar z a}]=0
\label{tre}
\ea
Using  $A_{\bar z}=0$ and   the gauge condition (\ref{gauge}), eq.(\ref{tre})
reads
\ba
&&
\partial_\rho b_z = 0
\n
&&
\partial_\rho b_{\bar z} =0
\n
&&
\partial_z b_{\bar z} - \partial_{\bar z} b_z + i[b_z, b_{\bar z}]
=0
\ea
One then sees that $b_z,b_{\bar z}$  must be independent of $\rho$. Being the
boundary condition $b_{\bar z}|_{\partial M}=0$, this implies
that $b_{\bar z}=0$ everywhere. The remaining equation is
\ba
&&\partial_{\bar z} b_z =0
\ea
and then $b_z=b_z(z)$.
With this solution for the $U(1)$ field, the eq.(\ref{u})
simplifies to
\be
\partial_{\bar z}A^a_{z}=0
\ee
this implying $A_z=A_z(z)$.

Then, the general solution to eqns.(\ref{eqs}) with boundary conditions
$A_{\bar z}|_{\partial M}=b_{\bar z}|_{\partial M} =0$,
closely related to the 3d black hole,
is chiral,
\begin{eqnarray}
 A_z &=& d^{-1}\tilde A_z(z) d \nonumber \\
 A_{\bar z}&=&0 \nonumber\\
 A_\rho&=& i J_3 = d^{-1}\partial_\rho d \nonumber\\
 b_z&=&b_z(z) \nonumber\\
 b_\rho &=& b_{\bar z} = 0
  \label{A01}
\end{eqnarray}
with $\tilde A_z(z), b_z(z)$  arbitrary Lie algebra-valued functions of $z$.
This
configuration solves both, the commutative and noncommutative equations.
It can
also be checked that it is a fixed point under the Seiberg-Witten map \cite{SW}.
A similar analysis can be done for the second complex field $\bar {\cal A}$
leading to a solution analogous to (\ref{A01}) but with $A_z(z) \to
\bar A_{\bar z}(\bar z)$, $b_z(z) \to
\bar b_{\bar z}(\bar z)$ and $d \to d^{-1}$.

A gauge transformation (with group element $d^{-1}$) brings the solution
to the simpler form
\begin{eqnarray}
 A_z &=& A_z(z) \nonumber \\
 A_{\bar z}&=& A_\rho=0 \nonumber\\
 b_z&=&b_z(z) \nonumber\\
 b_{\bar z}&=& b_\rho= 0
 \label{A0}
\end{eqnarray}

An important property of (\ref{A0}) is its Kac-Moody symmetry under holomorphic
gauge transformations. To see this, let us
specialize to the case $b_z = 0$ and note that
 the configuration (\ref{A0}) is form-invariant under
gauge transformations which only depend on $z$. Let $\lambda = \lambda(z)$. We act
with the noncommutative transformation (\ref{deltaA}) and find,
\begin{eqnarray}
\delta A_z &=& \pp_z \lambda + A_z*\lambda - \lambda * A_z= \pp_z \lambda +
A_z\lambda - \lambda A_z,\\
\delta A_{\bar z}&=& 0,\\
\delta A_\rho &=& 0,
\end{eqnarray}
The $*$ product has been eliminated
 because the whole solution only depends on $z$.
This symmetry of the space of solutions (\ref{A0}) is generated by a Kac-Moody
algebra and play an important role in various approaches to understand the 3d
black
hole entropy as well as the Brown-Henneaux conformal symmetry.

\subsection{The metric}

Let us construct the metric corresponding to the solution found above.
We start from eq.({\ref{g}) with  the vierbeins $e_\mu$   contructed according to
eq.(\ref{ewA}) which, for the affine solution  takes the form
\ba
e_z^a J_a =\frac {l} {2i} d^{-1}\tilde A(z)
d\;\;\;\;\;\;\;&\;&\;\;\;\;\;\;\;\;\;
e_{\bar z}^aJ_a = -\frac {l} {2i} d\,\tilde {\bar A}(\bar z)
\,d^{-1}
\n
e_\rho^a J_a &=& l J_3
\ea
Defining
\ba
\tilde A =
\frac{i}{2} \left(
\begin{array}{ll}
A^3&A^+\\
A^-&-A^3
\end{array}
\right)
\;\;\;\;\;\;\;\;\;\;\;\;
\tilde{\bar A} =
\frac{i}{2}\left(
\begin{array}{ll}
\bar A^3&\bar A^+\\
\bar A^-&-\bar A^3
\end{array}
\right)
\label{gtz}
\ea
then, the symmetric (arc length) part of the associated metric is $ds^2=g_{\mu\nu}
dx^\mu dx^\nu$,
\ba
ds^2&=& l^2 d\rho^2 - \frac {l^2}4\left({A^3}^2+ A^+A^-\right)dz^2
- \frac {l^2}4\left({\bar {A^3}}^2+ \bar A^+\bar A^-\right)d\bar z^2
\n&&+\;
\frac {l^2}8\left(2 \{A^3,\bar A^3\}_+ + \{A^-,\bar A^+\}_+
 e^{-2\rho}+ \{A^+,\bar
A^-\}_+ e^{2\rho}
\right)dzd\bar z
\n && +\;i l^2 \bar A^3 d\bar z d\rho
- i l^2 A^3 dzd\rho
\label{gtzb}
\ea

At this point, we are interested in determining the conditions   to be imposed on
the gauge fields in order to have an asymptotically AdS metric. To this end, we
follow \cite{CHvD} extended to the noncommutative case. The non-diagonal
components should be absent. This can be achieved taking $A^3=\bar A^3=0$,
conditions that extend to the noncommutative case  the first Polyakov reduction
condition. The resulting metric is
\ba
ds^2&=& l^2 d\rho^2 - \frac {l^2}4 A^+A^- dz^2
- \frac {l^2}4 \bar A^+\bar A^- d\bar z^2
\n&&+\;
\frac {l^2}8\left(\{\bar A^+,A^-\}_+ e^{-2\rho}+
\{A^+,\bar A^-\}_+
e^{2\rho}\right)dzd\bar z
\ea
which has an asymptotic ($\rho \to \infty$) form
\be
ds^2= l^2 d\rho^2
+\frac {l^2}8  \{A^+,\bar A^-\}_+
e^{2\rho}dzd\bar z
\ee
Then,
to match with the AdS form we need to impose the condition
\be
\{A^+,\bar A^-\}_+ = 8
\ee
Taking the derivatives with respect to $z$ and $\bar z$ we obtain
the relations (remember that $A^+$ is holomorphic and $\bar A^-$ is
antiholomorphic)

\be
\{\partial_z A^+,\bar A^-\}_+ = 0
\;\;\;\;\;\;\;\;\;\;\;\;\;\;
\{A^+, \partial_{\bar z}\bar A^-\}_+ = 0
\label{not}
\ee
In the usual commutative case these relations will imply constants $A^+, \bar
A^-$. To test this in the noncommutative case, let us first observe that
following  \cite{Moreno:2001kt}, one can write
\ba
f(z)*g(\bar z) &=& e^{\frac{\theta}{2} \partial\bar \partial} f(z)g(\bar z)
\n
g(\bar z)*f(z) &=& e^{-\frac{\theta}{2} \partial\bar \partial} f(z)g(\bar z)
\label{cito}
\ea
which implies
\ba
\frac 1 2 \left\{f,g\right\} &=& \frac 1 2 \left(
e^{\frac{\theta}{2} \partial\bar \partial}
+
 e^{-\frac{\theta}{2} \partial\bar \partial}
\right)
f(z) g(\bar z)
\n
&=& \cosh\left(\frac{\theta}{2}\partial\bar \partial\right)f(z)g(\bar z)
\label{pobre}
\ea
Using this, eqs.(\ref{not}) can be rewritten as
\be
\cosh\!\left({\frac{\theta}{2} \partial \bar \partial}\right)(\partial_z A^+\bar
A^-) = 0
\;\;\;\;\;\;\;\;\;\;\;\;\;\;
\cosh\!\left({\frac{\theta}{2} \partial \bar \partial}\right)
(A^+ \partial_{\bar z}\bar A^-
)=0
\label{larga}
\ee
Calling $\psi_\lambda$ and $\lambda$ the eigenfunctions and eigenvalues of
$\partial\bar \partial$ and assuming that $\{\psi_\lambda\}$ is complete,
one can write
$\cosh((\theta/2) \partial\bar \partial) = \sum_\lambda \cosh((\theta/2)
 \lambda)
|\psi_\lambda\rangle \langle \psi_\lambda|$
This ensures that $\cosh\!\left({(\theta/2) \partial
\bar \partial}\right)$ has
no zero modes and then one has, from (\ref{larga})
\be
\partial_z A^+\bar A^- = 0
\;\;\;\;\;\;\;\;\;\;\;\;\;\;
A^+ \partial_{\bar z}\bar A^-
=0
\ee
this implies that $A^+, \bar A^-$ should be constants. Then we
have found the second reduction condition
\be
A^+ = 2 \;\;\;\;\;\;\;\;\;\;\;\;\;\;\;\;\;\;\; \bar A^-=2
\ee

We conclude that in order to have an asymptotic AdS
form in the noncommutative case, one needs to impose
just the usual Polyakov reduction conditions, previously discussed in \cite{CHvD}.
In this case, eqns(\ref{gtz}) take the form
\ba
A_z&=&
i\left(
\begin{array}{ll}
0 & e^\rho \\
 \frac 1 {2l} T(z)e^{-\rho}& 0
\end{array}
\right)
\label{Aredd}\\
\bar A_{\bar z}&=&
i\left(
\begin{array}{ll}
0 &  \frac 1 {2l} \bar T(\bar z) e^{-\rho}\\
e^\rho& 0
\end{array}
\right)
\label{Areddbar}\\
A_\rho &=& -\bar A_\rho = i J_3
\n
A_{\bar z}&=&\bar A_z = b = \bar b =0
\ea
With this, the symmetric metric as defined in (\ref{gtzb}) becomes
\ba
ds^2 = l^2 d\rho^2 - \frac {l}2 T dz^2
- \frac {l}2 \bar T d\bar z^2 +
\frac {1}8\left(\{\bar T, T\}_+  e^{-2\rho}+
8 l^2
e^{2\rho}\right)dzd\bar z
\ea
We see that the only component of the symmetric metric affected by
noncommutativity is $g^S_{z\bar z}$.  Using (\ref{cito}), this component can be
written as
\be
g^S_{z\bar z} = \cosh({\theta\over 2}\partial{\bar\partial}){\tilde g}_{z \bar z}
\ee
being $\tilde g$ the metric constructed in \cite{Baires} for
the commutative case.  The operator $\cosh((\theta/2)\partial\bar \partial)$ acts
like the identity when applied to the other components of the
metric (all derivative terms vanishes),
\ba
g^S_{zz} &=& \cosh(\frac{\theta}{2}\partial\bar \partial) \tilde g_{zz}
\n
g^S_{\bar z \bar z} &=& \cosh(\frac{\theta}{2}\partial\bar \partial) \tilde
g_{\bar
z\bar z}
\ea
so that the relation between the commutative and the (symmetric) noncommutative
solutions can be compactly written as
\be
g^S_{\mu\nu}= \cosh\left(\frac{\theta}{2}\partial\bar \partial\right)
\tilde g_{\mu\nu}.
\ee

The full metric $g_{\mu\nu} = g^S_{\mu\nu} + g^A_{\mu\nu}$, where $g^A_{\mu\nu}$
is the anti-symmetric part, satisfies the ``Einstein" equation (\ref{EE}). Note
that $g^A_{\mu\nu}$ is in fact non-zero.  Its non-zero contributions come from
\be
g_{z\bar z} = \exp\left(\frac{\theta}{2}\partial\bar \partial\right)
{\tilde g_{z\bar z}}  \, ,  \;\;\; \;\;\; \;\;\;
g_{\bar z z} = \exp\left(-\frac{\theta}{2}\partial\bar \partial\right)
{\tilde g_{\bar z z}}
\ee
which imply
\begin{equation}
g^A_{z \bar z} = \sinh\left(\frac{\theta}{2}\partial\bar \partial\right)
{\tilde g_{z\bar z}}
\label{}
\end{equation}

Recall that the deviation of (\ref{EE}) from the ordinary Einstein equations is
encoded in the combination ${E^\mu} _{\nu\alpha\beta}$ which depends on the
commutator $[{e^a} _\alpha,{e^b}_\beta]$.
In the present case the only non-vanishing contribution to this commutator is the
$(\alpha=z,\beta={\bar z})$ component, and it is proportional to the commutator
$[T,{\bar T}] = 2 \sinh \left(\frac{\theta}{2}\partial\bar \partial\right) T(z)
\bar T(\bar z)$.

For future use, let us end this section rewriting the solution
(\ref{Aredd})-(\ref{Areddbar}) in the $A_\rho = 0$ gauge
\begin{equation}
A_z = i\left( \begin{array}{cc}   0 &  1  \\
                           {1 \over 2l}T(z) &  0   \end{array} \right)
\label{Ared}
\end{equation}
\begin{equation}
\bar A_{\bar z} = i\left( \begin{array}{cc}   0 &  {1 \over 2l} \bar T(\bar z)
 \\
                           1 &  0   \end{array} \right)
\label{Aredbar}
\end{equation}


\section{Constant Abelian background}

We  consider in this section the chiral solution considered in the last section
coupled to a constant Abelian field of magnitude $F_{z\bar z}= i\alpha$.  We shall
see that the black hole field with constant values of $T$ and $\bar T$ will feel
the
Abelian field due to noncommutative effects.

In order to fix the value of the Abelian field we
add to the action the term
$-2i\int \mbox{Tr} (\alpha \, {\cal A})$
where $\alpha$ is a fixed 2-form $\alpha=\alpha
dz\ww d\bar z$. This is a term of the kind introduced in \cite{Polych}.   The
equations of motion (\ref{F=0}) are replaced by
\begin{equation}
\pp_\mu  {\cal A}_\nu - \pp_\nu  {\cal A}_\mu + {\cal A}_\mu *
{\cal A}_\nu - {\cal A}_\nu *
{\cal A}_\mu = \alpha_{\mu\nu}\, i  .
\label{F=alpha}
\end{equation}
$\alpha$ is a number and it
contributes only to the Abelian curvature\footnote{A
constant noncommutative Abelian field has been studied in detail in
\cite{AlvarezG}}.

The generalization of the chiral solution satisfying (\ref{F=alpha})
in the ${\cal A}_\rho = 0$ gauge is simply,
\begin{eqnarray}
 {\cal A}_z &=& A(z) - i\,\alpha \, \bar z \nonumber  \\
 {\cal A}_{\bar z}&=& 0   \label{A2} \\
 {\cal A}_\rho&=& 0. \nonumber
\end{eqnarray}
Since the extra term only contributes to the Abelian field, one could naively
conclude that the black hole solution has not changed.  However, this field
depends
on both coordinates and noncommutative effects do take place.

The point is that, the noncommutative structure of the gauge transformations
changes the affine algebra and, as a result, Polyakov's reduction conditions needs
to be modified.  Let $\lambda=\lambda(z)$ and compute the noncommutative gauge
transformation (\ref{deltaA}) acting on (\ref{A2}). The components
${\cal A}_\rho$ and $
{ \cal A}_{\bar z}$ are left invariant while the transformation for
$ {\cal A}_z$ yields,
\begin{eqnarray}
\delta  {\cal A}_z&=& \pp_z \lambda + (A_z-i\alpha \,\bar  z) *\lambda - \lambda *
(A_z-
i\alpha
\,\bar z) , \nonumber \\
          &=& (1 +i \theta \, \alpha) \pp_z \lambda +  A_z\lambda - \lambda
          A_z \label{dA}
\end{eqnarray}
The extra term proportional to $\theta$ comes from the Moyal formula $\bar z
* f - f*\bar z = -\theta \,\pp f$.  The solution (\ref{A2}) still has an affine
holomorphic Kac-Moody symmetry but its form has changed.

Even though the extra term $\alpha\theta$ in (\ref{dA})  does not affect the gauge
symmetries in any significant way\footnote{In fact one could define $A=
(1+\alpha\theta) A'$ and $A'$ would transform in the usual way. This corresponds
to
the Seiberg-Witten map \cite{SW} applied to this particular situation.},
it does change the definition of global charges. We shall see that the mass and
angular momentum of
the black hole are modified by the presence of $\alpha$.

The point is that under the transformation (\ref{dA}), the reduction condition
$A^+_z= 2$ is not consistent, and does not yield the Virasoro algebra. The correct
reduction conditions are
\begin{equation}
A^3_z=0, \ \ \ \  A^+_z=2(1 + i \alpha \, \theta),
\label{red1}
\end{equation}
and the Virasoro charge is $T(z) = A^-/(2+2i \alpha\theta)$. The
reduced field is then
\begin{equation}
A_z = i(1+i \alpha\theta)\left( \begin{array}{cc}   0 &  1  \\
                              {T(z)\over 2l} &  0   \end{array}
                               \right)
\label{Ared1}
\end{equation}
In order to match the boundary conditions (keeping the periodicity of the torus
fixed) with the solution (\ref{Ared}) we perform a
constant gauge transformation on $A_z$ with a group element $g=e^{i a J_3}$ and
$a=
\log(1+i\alpha\theta)$. The field  (\ref{Ared1}) is transformed into
\begin{equation}
A_z = i\left( \begin{array}{cc}   0 & 1   \\
                               (1+ i\alpha\theta)^2 {T(z)\over 2l} &  0
                               \end{array}
                               \right)
\label{Ared2}
\end{equation}
which is of the form (\ref{Ared}). The antiholomorphic  field can be constructed
in
a
similar way and one finds,
\begin{equation}
\bar A_{\bar z} = i\left( \begin{array}{cc}   0 &  (1-i\bar \alpha\theta)^2 {\bar
T(\bar z)
\over 2l}   \\
                               1 &  0   \end{array}
                               \right)
\label{Ared2bar}
\end{equation}

For constant values of $T$ and $\bar T$ this field represent a black hole. However
the relation between the mass and angular momentum and the Virasoro charges have
changed, \begin{eqnarray}
Ml &=& (1+i\alpha \theta)^2 T + (1-i \bar \alpha \theta)^2 \bar T, \\
i J  &=& (1+i\alpha \theta)^2 T - (1-i \bar \alpha \theta)^2 \bar T,
\label{MJ2}
\end{eqnarray}

It is instructive to expand these relations to first order in $\theta$,
\begin{eqnarray}
l M &=& l M_0 + 2 \theta \alpha \, J_0  \\
J  &=& J_0 - 2 \theta \alpha \, l M_0,
\label{MJ3}
\end{eqnarray}
where $M_0$ and $J_0$ are the values of $M$ and $J$ at $\alpha=0$.  For
example one can start at $\alpha=0$ with a non-rotating black hole ($J_0=0$).
Then we turn on the Abelian field with $\alpha \neq 0$ and
find that the corresponding black hole will have a non-zero angular momentum.

\vspace{1 cm}

\noindent\underline{Acknowledgements}: M.B. is supported by Grant \#
1000744 (FONDECYT, Chile) and O.C. by Grant \# 3000026  (FONDECYT, Chile). N.G.,
F.A.S and G.A.S. are partially supported by UNLP, CICBA, CONICET (PIP 4330/96),
ANPCYT (PICT 03-05179), Argentina. M.B. would like to thank Marc Henneaux for a
useful conversation on noncommutative field theories.

\end{document}